\documentclass[preprint,12pt]{elsarticle}
\usepackage{amssymb,amsmath,amscd}
\usepackage{amsfonts,bm,latexsym}

\journal{Physics Letters B}

\def\be{\begin{equation}}
\def\ee{\end{equation}}
\def\bea{\begin{eqnarray}}
\def\eea{\end{eqnarray}}

\def\nn{\nonumber}
\def\p{\partial}
\def\oneone{\rlap 1\mkern4mu{\rm l}}
\def\Star{\,^{\star}\!}

\begin{document}
\begin{frontmatter}

\title{Two-charged non-extremal rotating black holes in seven-dimensional gauged supergravity:
The single-rotation case}

\author{Shuang-Qing Wu} 
\address{College of Physics and Electric Information, China West Normal University,
Nanchong, Sichuan 637009, People's Republic of China}

\begin{abstract}
We construct the solution for non-extremal charged rotating black holes in seven-dimensional
gauged supergravity, in the case with only one rotation parameter and two independent charges.
Using the Boyer-Lindquist coordinates, the metric is expressed in a generalized form of the
ansatz previously presented in [S.Q. Wu, Phys. Rev. D \textbf{83} (2011) 121502(R)], which
may be helpful to find the most general non-extremal two-charged rotating black hole with
three unequal rotation parameters. The conserved charges for thermodynamics are also computed.
\end{abstract}


\end{frontmatter}

\section{Introduction}

The discovery of the AdS/CFT correspondence stimulates a great deal of interest in constructing
rotating charged black holes in gauged supergravities during the last few years. Of particular
interest are those non-extremal black hole solutions in spacetime dimensions $D = 4, 5, 7$,
which respectively correspond to the maximal $D = 4$, $\mathcal{N} = 8$, SO(8); $D = 5$,
$\mathcal{N} = 8$, SO(6); and $D = 7$, $\mathcal{N} = 4$, SO(5) gauged supergravities with
respective Cartan subgroups U$(1)^4$, U$(1)^3$ and U$(1)^2$. In recent years, there has been
much progress in obtaining new, non-extremal, asymptotically AdS black hole solutions of gauged
supergravity theories in four, five, six and seven dimensions. For a comprehensive discussion
of these solutions, see for example \cite{chow}. However, almost all of these solutions can be
classified into three catalogues: either setting all of the angular momenta equal; or setting
certain U(1) charges equal; or restricting to supersymmetric solutions. One main reason is that,
for non-extremal solutions of gauged supergravity theories, there is no known solution generating
technique that can charge up a neutral solution, instead one must rely on inspired guesswork and
apply these strategies to greatly simplify the problem in finding exact solutions. So far, there
is no universal method to derive the above-mentioned solutions, namely those listed in \cite{chow}.

In a recent paper \cite{SQWu} that extends an interesting work \cite{Chow10} in $D = 4$ to all
dimensions, we have presented in a unified fashion the general non-extremal
rotating, charged Kaluza-Klein AdS black holes with only one electric charge and with arbitrary
angular momenta in all higher dimensions. The solutions in $D = 4, 5, 6, 7$ dimensions can be
embedded into corresponding gauged supergravities and the scalar potential in the Lagrangian
can be rigorously deduced via the Killing spinor equation \cite{LLW}. What is more, it has been shown
that the general solutions in all dimensions share a common and universal metric structure which
can not only naturally reduce to the famous Kerr-Schild ansatz in the uncharged case, but also be
generalized to all of already-known black hole solutions with multiple pure electric charges, both
in the cases of rotating charged black holes in ungauged supergravity and in the cases of nonrotating
AdS black holes in gauged supergravity theories. This means that all previously-known supergravity
black hole solutions with multiple different electric charges can be recast into a unified metric
ansatz, regardless they belong to ungauged theories or gauged ones. In other words, supergravity
black hole solutions in gauged theory inherit the same underlying metric structure as their ungauged
counterparts. This significant feature of supergravity black hole solutions had not been exploited
in any other previous work. Although that work \cite{SQWu} only dealt with the single-charge case
in Kaluza-Klein supergravity, it put forward a universal method to construct the most general rotating
charged AdS black hole solutions with multiple pure electric charges in gauged supergravity theory.
For example, guided by the generalized form of that ansatz, the most general charged rotating AdS$_5$
solution with three unequal charges and with two independent rotation parameters has been successfully
constructed \cite{Wu} within five-dimensional $U(1)^3$ gauged supergravity.

In this Letter, we shall concentrate on constructing new non-extremal rotating charged AdS black
hole solutions in the dimension $D = 7$ which are relevant for the AdS$_7$/CFT$_6$ correspondence
in M-theory. The $D = 7$ case singles it out a unique role since there appears a first-order
``odd-dimensional self-duality'' for the 4-form field strength, not seen previously in lower-dimensional
examples. This feature makes it quite complicated for finding an exact solution. The currently-known
charged non-extremal black hole solutions in the seven-dimensional gauged supergravity are as follows.
Non-rotating static charged AdS$_7$ black hole solutions were known \cite{MCetal,CG,LM} for a
decade because of the AdS/CFT correspondence. The first non-extremal rotating charged AdS black
holes with two different electric charges in seven-dimensional gauged supergravity were obtained
in \cite{CCLP}, in the special case where the three rotation parameters are set equal. Inspired by
a special case \cite{CCLP} where both U(1) charges are set equal, Chow \cite{chow08} found a new
rotating charged AdS black hole solution with three independent angular momenta and two equal U(1)
charges. The single-charge case within $D = 7$ Kaluza-Klein supergravity theory was recently found
in \cite{SQWu}. However, the most general non-extremal rotating charged AdS black hole solution with
all three unequal rotation parameter and two different U(1) charges is still not yet known.

In the present Letter, we shall construct new solution for non-extremal charged rotating black holes
in seven-dimensional gauged supergravity, in the case with two independent charges but with only one
rotation parameter. This is helped by the metric structure found for the two-charged Cveti\v{c}-Youm
solution \cite{CY2c} in arbitrary dimensions, which is presented in Appendix A. It should also be
pointed out that the recently-found two-charge rotating black holes \cite{Chow11} in four-dimensional
gauged supergravity can be cast into the same ansatz. We present the solution in Section 2 and then
examine its thermodynamics in Section 3. The conclusion section and the remaining two appendices B
and C summarize our results and present the clue toward constructing the most general solution with
two unequal charges and with three unequal rotation parameters in seven-dimensional gauged supergravity.

\section{Non-extremal rotating charged
solution with one angular momentum}

The $\mathcal{N} = 4$ seven-dimensional gauged supergravity theory is a consistent reduction
of eleven-dimensional supergravity on $S^4$. It is capable of supporting black holes with two
independent electric charges, carried by gauge fields in the $U(1)\times U(1)$ Abelian subgroup
of the full $SO(5)$ gauge group. After a further consistent truncation, in which all except
the $U(1)\times U(1)$ subgroup of gauge fields are set to zero, the fields in the final theory
comprise the metric, two dilatons, two $U(1)$ gauge fields and a 4-form field strength that
satisfies an odd-dimensional self-duality equation. For the purpose of this Letter, we will
consider the bosonic sector of the seven-dimensional gauged supergravity that consist of a
graviton, two dilaton scalars, two Abelian gauge potentials and a 3-form potential, whose
Lagrangian is given by \cite{CCLP}
\bea
\mathcal{L} &=& R \star\oneone -\Star d\varphi_1\wedge d\varphi_1 -5\Star d\varphi_2\wedge d\varphi_2
 -\frac{1}{2}X_1^{-2}\Star F_1\wedge F_1 -\frac{1}{2}X_2^{-2}\Star F_2\wedge F_2 \nn \\
&& -\frac{1}{2}(X_1X_2)^2\Star\mathcal{F}\wedge \mathcal{F} +(F_1\wedge F_2 -g\mathcal{F})
 \wedge\mathcal{C} \nn \\
&& +2g^2\big[8X_1X_2 +4X_1^{-1}X_2^{-2} +4X_1^{-2}X_2^{-1} -(X_1X_2)^{-4}\big] \star\oneone \, , \label{d7L}
\eea
where
\be
F_1 = dA_1 \, , \quad F_2 = dA_2 \, ,\quad \mathcal{F} = d\mathcal{C} \, , \qquad
X_1 = e^{-\varphi_1 -\varphi_2} \, ,\quad X_2 = e^{\varphi_1 -\varphi_2} \, ,
\ee
together with a first-order odd-dimensional self-duality equation
\be
(X_1X_2)^2\Star\mathcal{F} = -2g\mathcal{C} -\mathcal{H} \, , \label{sde}
\ee
satisfied by the 4-form field strength, which is conveniently stated by introducing an
additional 2-form potential $\mathcal{B}$
\be
\mathcal{H} = d\mathcal{B} -(A_1\wedge F_2 +A_2\wedge F_1)/2 \, .
\ee

Now we present the exact solution for non-extremal charged rotating black holes in the above
theory, in the case with only one rotation parameter and with two independent charges. In terms
of the generalized Boyer-Lindquist coordinates, the metric is written in a generalized form
of the ansatz given in a previous work \cite{SQWu}, which sheds light on how to find the most
general non-extremal two-charged rotating black hole with three unequal rotation parameters.
The metric and two $U(1)$ Abelian gauge potentials have the following exquisite form
\bea
&&ds^2 = (H_1H_2)^{1/5}\bigg[ -\frac{(1+g^2r^2)\Delta_{\theta}}{\chi} dt^2
 +\frac{\Sigma}{\Delta_r} dr^2 +\frac{\Sigma}{\Delta_{\theta}} d\theta^2 \nn \\
&&\qquad +\frac{(r^2+a^2)\sin^2\theta}{\chi} d\phi^2
 +r^2\cos^2\theta\big(d\psi^2 +\cos^2\psi d\zeta^2 +\sin^2\psi d\xi^2\big) \nn \\
&&\qquad +\frac{2ms_1^2}{r^2\Sigma H_1\chi^2(s_1^2-s_2^2)}k_1^2
 +\frac{2ms_2^2}{r^2\Sigma H_2\chi^2(s_2^2-s_1^2)}k_2^2 \bigg] \, , \\
&&A_i = \frac{2ms_i}{r^2\Sigma H_i\chi}k_i \, ,
\eea
in which
\be
k_1 = c_1\sqrt{\Xi_2}\Delta_{\theta} dt -c_2\sqrt{\Xi_1}a\sin^2\theta d\phi \, , \quad
k_2 = c_2\sqrt{\Xi_1}\Delta_{\theta} dt -c_1\sqrt{\Xi_2}a\sin^2\theta d\phi \, ,
\ee
and
\bea
\Delta_r &=& \big(r^2 +a^2 -2m/r^2\big)\big(1 +g^2r^2 -2mg^2s_1^2s_2^2/r^2\big)
 +2mg^2c_1^2c_2^2 \, , \nn \\
\Delta_{\theta} &=& 1 -g^2a^2\cos^2\theta \, , \qquad \Sigma = r^2 +a^2\cos^2\theta \, , \nn \\
H_i &=& 1 +2ms_i^2/(r^2\Sigma) \, , \qquad \Xi_i = c_i^2 -s_i^2\chi \, , \quad
\chi = 1 -g^2a^2 \, , \nn
\eea
where the short notations $c_i = \cosh\delta_i$ and $s_i = \sinh\delta_i$ ($i=1,2$) are used.

Two scalars are given by $X_i = (H_1H_2)^{2/5}/H_i$, while the non-vanishing components of the
2-form potential and 3-form potential are
\bea
\mathcal{B}_{t\phi} &=& \frac{ms_1s_2\Delta_{\theta}a\sin^2\theta}{r^2\Sigma\chi}\Big(\frac{1}{H_1}
 +\frac{1}{H_2}\Big) \, , \\
\mathcal{C}_{t\theta\phi} &=& g\frac{2ms_1s_2a\sin\theta\cos\theta}{r^2\chi} \, , \quad
\mathcal{C}_{\psi\zeta\xi} = \frac{2ms_1s_2a\cos^4\theta}{\Sigma}\sin\psi\cos\psi \, .
\eea

We have found this solution by firstly recasting the metric and two gauge fields into a
generalized ansatz in which two vectors $k_1$ and $k_2$ can be easily written down. This
solution ansatz is inspired from our observation that the general two-charged rotating
solutions \cite{CY2c} in all dimensions in ungauged supergravity and a special two-charged
rotating AdS$_7$ gauged supergravity solution \cite{CCLP} with equal rotation parameters
can be recast into a similar form, which are respectively rewritten in Appendix A and B.
Next, we can decide the radial function $\Delta_r$ by requiring the metric determinant
to be the expected expression. After doing this, we further find the 3-form potential
$\mathcal{C}$ via solving the self-duality equation (\ref{sde}), since the expression
for the 2-form potential $\mathcal{B}$ can be easily conjectured according to our previous
experience. The last thing is to mechanically verify that the solution indeed solves all
the field equations derived from the Lagrangian (\ref{d7L}).

\section{Thermodynamics}

The two-charged AdS$_7$ black holes have Killing horizons at $r = r_+$, the largest positive
root of $\Delta_r = 0$. The entropy and the Hawking temperature of the outer horizon are
easily evaluated as
\bea
S &=& \frac{\pi^3r_+}{4\chi}\sqrt{r_+^2(r_+^2 +a^2) +2ms_1^2}\sqrt{r_+^2(r_+^2 +a^2) +2ms_2^2} \, , \\
T &=& \frac{r_+^2\Delta^{\prime}_{r_+}}{4\pi\sqrt{r_+^2(r_+^2 +a^2)
 +2ms_1^2}\sqrt{r_+^2(r_+^2 +a^2) +2ms_2^2}} \, .
\eea

On the horizon, the angular velocity and the electrostatic potentials are given by
\bea
\Omega &=& \frac{2mr_+^2ac_1c_2\sqrt{\Xi_1\Xi_2}}{\big[r_+^2(r_+^2 +a^2)
 +2ms_1^2\big]\big[r_+^2(r_+^2 +a^2) +2ms_2^2\big]} \nn \\
&=& \frac{a}{2mr_+^2c_1c_2\sqrt{\Xi_1\Xi_2}}\big(r_+^2 +g^2r_+^4 +2mg^2s_1^2\big)
 \big(r_+^2 +g^2r_+^4 +2mg^2s_2^2\big) \, , \quad \\
\Phi_1 &=& \frac{2mc_1s_1\sqrt{\Xi_2}}{r_+^2(r_+^2 +a^2) +2ms_1^2} \, , \qquad
\Phi_2 = \frac{2mc_2s_2\sqrt{\Xi_1}}{r_+^2(r_+^2 +a^2) +2ms_2^2} \, .
\eea

The metric is obviously asymptotic anti-de Sitter, at infinity it approaches to the
conformal metric
\bea
\lim_{r\to\infty}\frac{ds^2}{r^2} &=& -\frac{g^2\Delta_{\theta}}{\chi}\, dt^2
 +\frac{dr^2}{g^2r^4} +\frac{d\theta^2}{\Delta_{\theta}}
 +\frac{\sin^2\theta}{\chi}\, d\phi^2 \nn \\
 && +\cos^2\theta\big(d\psi^2 +\cos^2\psi d\zeta^2 +\sin^2\psi d\xi^2\big) \, ,
\eea
with which we can choose two vectors
\bea
\hat{N}^a = \frac{\sqrt{\chi}}{g\sqrt{\Delta_{\theta}}}(\p_t)^a \, , \qquad
\hat{n}^a = -g^2r^2(\p_r)^a \, , \nn
\eea
to compute the conserved charges that obey thermodynamical first laws.

Using the formulae \cite{CLP06}
\bea
\mathcal{Q}[\xi] &=& \frac{1}{32\pi}\int_{S^5} d^5x \frac{\sin\theta\cos^3\theta\sin\psi
 \cos\psi}{g^3\sqrt{\chi\Delta_{\theta}}} r^2C_{acbd}\xi^{a}\hat{N}^b\hat{n}^c\hat{n}^d \nn \\
&=& \frac{\pi^2}{8}\int_0^{\pi/2} d\theta \frac{\sin\theta\cos^3\theta}{\Delta_{\theta}}
 r^6C_{artr}\xi^{a} \, ,
\eea
one can compute the conserved mass and angular momentum as
\bea
M &=& -\mathcal{Q}[\p_t] = \frac{\pi^2 m}{8\chi}\bigg[2c_1^2c_2^2
 \Big( 1 +\frac{1}{\chi}\Big) +1 -s_1^2s_2^2(1 +3\chi)\bigg] \, , \\
J &=& \mathcal{Q}[\p_{\phi}] = \frac{\pi^2 mac_1c_2\sqrt{\Xi_1\Xi_2}}{4\chi^2} \, ,
\eea
while two electric charges can be easily evaluated as
\bea
Q_1 &=& \frac{1}{16\pi}\int_{S^5} \big(X_1^{-2}\Star F_1 -F_2\wedge \mathcal{C})
 = \frac{\pi^2 mc_1s_1\sqrt{\Xi_2}}{2\chi} \, , \\
Q_2 &=& \frac{1}{16\pi}\int_{S^5} \big(X_2^{-2}\Star F_2 -F_1\wedge \mathcal{C})
 = \frac{\pi^2 mc_2s_2\sqrt{\Xi_1}}{2\chi} \, .
\eea

These thermodynamical quantities satisfy the differential and integral first laws of
thermodynamics
\bea
&& dM = T dS +\Omega dJ +\Phi_1 dQ_1 +\Phi_2 dQ_2 -P d\mathcal{V} \, , \\
&& \frac{4}{5}(M -\Phi_1 Q_1 -\Phi_2 Q_2) = T\, S +\Omega J -P\, \mathcal{V} \, ,
\eea
where we have introduced the generalized pressure \cite{CCK}
\bea
P &=& \frac{g^5m}{20\pi\chi}\bigg\{c_1^2c_2^2\bigg[\frac{1}{\chi} -1
 +\frac{g^2r_+^2(5r_+^2-8mg^2s_1^2s_2^2)}{r_+^2(1+g^2r_+^2) -2mg^2s_1^2s_2^2}\bigg] \nn \\
&&\qquad -2g^2r_+^2(s_1^2+s_2^2+2s_1^2s_2^2) \bigg\} \, ,
\eea
which is conjugate to the volume $\mathcal{V} = \pi^3/g^5$ of the $5$-sphere with the
AdS radius $1/g$. The results presented above include that given in Ref. \cite{SQWu}
as a special case when one charge and two rotation parameters are set to zero in $D = 7$
dimensions.

\section{Conclusions}

In this Letter, we have successfully constructed the non-extremal two-charged single-rotating
black holes in seven-dimensional gauged supergravity and computed their conserved charges
closely related to thermodynamical first laws. Our results are significant for testing the
AdS$_7$/CFT$_6$ correspondence in M-theory. To find this exact two-charged solution, we are
helped by generalizing the solution ansatz previously proposed in Ref. \cite{SQWu} to the case
with two different electric charges. Along the same line, it seems a direct and simple thing
to apply this ansatz to construct the most general non-extremal rotating charged black hole
solutions with two different charges and three unequal rotation parameters in seven-dimensional
gauged supergravity theory. However, a peculiar self-duality equation (\ref{sde}) in $D = 7$
makes the thing quite complicated because we must also simultaneously deal with the 2-form
potential and the 3-form potential. What is more, peered from the expressions of two $U(1)$
Abelian gauge potentials (\ref{tgp1},\ref{tgp2}) given in the Appendix B for the case with three
equal rotation parameters, we infer that the expressions for two $U(1)$ gauge potentials may
not be so simple to be conjectured for the expected solutions. Nevertheless, the ansatz
presented here already sheds new light on how to construct the most general rotating charged
black hole solutions with multiple different electric charges in gauged supergravity theory.
It deserves more deep investigations of the solution ansatz in the future work.

In addition, the corresponding supersymmetric solution, the separability of Hamilton-Jacobi
equation, massless Klein-Gordon equation and their related hidden symmetry of the solution
remain to be further investigated.

\textbf{Note added}: After our work appeared, the solution was also independently found by
using a non-universal ansatz and its properties were examined in Ref. \cite{Chow1108}.

\section*{Acknowledgements}
S.-Q. Wu is supported by the NSFC under Grant Nos. 10975058 and 10675051. The computation
within this work has been done by using the GRTensor-II program based on Maple 7. He is
grateful to Prof. Hong L\"{u} for useful discussions.

\appendix

\section{Two-charged Cveti\v{c}-Youm
solution in arbitrary dimensions}

The general non-extremal rotating charged black holes, with two different charges and with
arbitrary angular momenta in all higher dimensions, within ungauged supergravity theory were
obtained in Ref. \cite{CY2c} (see also \cite{HS,PML}) via solution-generating technique. These
solutions are relevant for toroidally compactified heterotic supergravity. Although we focus
here on the special case for $D = 7$, it is universal to recast the metric into an exquisite
ansatz in all dimensions. The underlying metric structure of the solutions can be easily
recognized as follows
\bea
ds^2 &=& (H_1H_2)^{\frac{1}{D-2}}\Big[ -dt^2
 +\sum_{i=1}^{N +\epsilon}(r^2+a_i^2) d\mu_i^2 +\sum_{i=1}^N(r^2+a_i^2)\mu_i^2 d\phi_i^2 \nn \\
&& +\frac{U}{V -2m} dr^2 +\frac{2ms_1^2}{UH_1(s_1^2-s_2^2)}\Big(c_1 dt
 -c_2\sum_{i=1}^N a_i\mu_i^2 d\phi_i\Big)^2 \nn \\
&&\quad +\frac{2ms_2^2}{UH_2(s_2^2-s_1^2)}\Big(c_2 dt
 -c_1\sum_{i=1}^N a_i\mu_i^2 d\phi_i\Big)^2 \Big] \, , \\
A_1 &=& \frac{2ms_1}{UH_1}\Big(c_1 dt -c_2\sum_{i=1}^N a_i\mu_i^2 d\phi_i\Big) \, , \\
A_2 &=& \frac{2ms_2}{UH_2}\Big(c_2 dt -c_1\sum_{i=1}^N a_i\mu_i^2 d\phi_i\Big) \, , \\
\mathcal{B} &=& -\frac{ms_1s_2}{U}\Big(\frac{1}{H_1} +\frac{1}{H_2}\Big)
 dt\wedge \sum_{i=1}^N a_i\mu_i^2 d\phi_i \, ,
\eea
where $H_i = 1 +2ms_i^2/U$ and
\bea
U = r^{\epsilon}\sum_{i=1}^{N +\epsilon}\frac{\mu_i^2}{r^2+a_i^2}\prod_{j=1}^N (r^2+a_j^2) \, , \qquad
V = r^{\epsilon-2}\prod_{i=1}^N(r^2+a_i^2) \, . \nn
\eea
In the above, we have denoted the dimension of spacetime as $D = 2N +1 +\epsilon \geq 4$, with
$N = [(D-1)/2]$ being the number of rotation parameters $a_i$ associated with the $N$ azimuthal
angles $\phi_i$ in the $N$ orthogonal spatial 2-planes, and $2\epsilon = 1 +(-1)^D$. The $N
+\epsilon = [D/2]$ `direction cosines' $\mu_i$ obey the constraint $\sum_{i=1}^{N +\epsilon}
\mu_i^2 = 1$, where $0\leq \mu_i \leq 1$ for $1\leq i\leq N$, and $-1\leq \mu_{N+1}\leq 1$
for even $D$.

\section{Two-charged solution
\cite{CCLP} in $D = 7$ dimensions}

The two-charged seven-dimensional rotating black hole solution \cite{CCLP}, with the three
rotation parameters being set equal, can be rewritten as
\bea
ds_7^2 &=& (H_1H_2)^{1/5}\bigg[ -\frac{1+g^2r^2}{\chi\Xi_-^2} dt^2 +\frac{(r^2+a^2)^2r^2}{Y}
 dr^2 +\frac{r^2+a^2}{\chi} d\Sigma_2^2 \nn \\
&& +\frac{r^2+a^2}{\chi}\Big(\sigma +\frac{g}{\Xi_-} dt\Big)^2
 +\frac{2ms_1^2}{(s_1^2-s_2^2)(r^2+a^2)^2H_1\chi^2}(\alpha_1 dt -\alpha_2 a\sigma)^2 \nn \\
&&\quad +\frac{2ms_2^2}{(s_2^2-s_1^2)(r^2+a^2)^2H_2\chi^2}(\alpha_2 dt -\alpha_1 a\sigma)^2 \bigg] \, , \\
A_1 &=& \frac{2ms_1}{(r^2+a^2)^2H_1\chi}(\alpha_1 dt -\alpha_2 a\sigma) \, , \label{tgp1} \\
A_2 &=& \frac{2ms_2}{(r^2+a^2)^2H_2\chi}(\alpha_2 dt -\alpha_1 a\sigma) \, , \label{tgp2} \\
\mathcal{B} &=& \frac{ms_1s_2}{(r^2+a^2)^2\Xi_-^2}\Big(\frac{1}{H_1} +\frac{1}{H_2}\Big)
 dt\wedge a\sigma \, , \quad
\mathcal{C} = \frac{mas_1s_2}{(r^2+a^2)\chi\Xi_-}\sigma\wedge d\sigma \, , \qquad
\eea
where
\bea
&& \alpha_1 = c_1 -(1-\Xi_+^2)(c_1 -c_2)/2 \, , \quad
\alpha_2 = c_2 -(1-\Xi_+^2)(c_2 -c_1)/2 \, , \nn \\
&& \Xi_\pm = 1\pm ga \, , \quad \chi = 1 -g^2a^2 \, . \nn
\eea
The two scalars are $X_i = (H_1H_2)^{2/5}H_i^{-1}$ with $H_i = 1 +2ms_i^2/(r^2+a^2)^2$, while
the function $Y$ is presented in Ref. \cite{CCLP} and is not given here. The unit 5-sphere is
parameterized by the Fubini-Study metric on $\mathbb{CP}^2$, and the connection on the $U(1)$
fibre over $\mathbb{CP}^2$
\bea
d\Sigma_2^2 &=& d\xi^2 +\frac{1}{4}\sin^2\xi(d\theta^2 +\sin^2\theta d\phi^2)
 +\frac{1}{4}\sin^2\xi\cos^2\xi(d\psi +\cos\theta d\phi)^2 \, , \nn \\
\sigma &=& d\tau +\frac{1}{2}\sin^2\xi(d\psi +\cos\theta d\phi) \, . \nn
\eea
When the rotation parameter is set to zero, the ansatz presented above contains as a special
case the nonrotating charged AdS$_7$ black hole solution found in Refs. \cite{MCetal,CG,LM}.

It should be pointed out that although the above solution can be recast into the general ansatz,
the forms for the two gauge potentials are rather complicated after make a re-scaling $t\to
\Xi_- \bar{t}$ and a shift $\sigma\to \bar{\sigma} -gd\bar{t}$. Because the metric contains
the similar expressions to those of two gauge potentials, thus it provides little useful
information on how to generalize the solution to the general case with three unequal rotation
parameters.

\section{A different but simpler
form for two-charged solution}

In this appendix, we present a different but simpler form for the two-charged seven-dimensional
rotating black hole solution with the three rotation parameters being set equal. Written in
terms of the previous ansatz, it is given by
\bea
ds_7^2 &=& (H_1H_2)^{1/5}\bigg[ -\frac{1+g^2r^2}{\chi} dt^2 +\frac{(r^2+a^2)^2r^2}{\Delta_r}
 dr^2 +\frac{r^2+a^2}{\chi} d\Sigma_2^2 \nn \\
&& +\frac{r^2+a^2}{\chi}\big(\sigma -g dt\big)^2
 +\frac{2ms_1^2}{(s_1^2-s_2^2)(r^2+a^2)^2H_1\chi^2}(\alpha dt -\beta\gamma a\sigma)^2 \nn \\
&&\quad +\frac{2ms_2^2}{(s_2^2-s_1^2)(r^2+a^2)^2H_2\chi^2}(\beta dt -\alpha\gamma a\sigma)^2 \bigg] \, , \\
A_1 &=& \frac{2ms_1}{(r^2+a^2)^2H_1\chi}(\alpha dt -\beta\gamma a\sigma) \, , \\
A_2 &=& \frac{2ms_2}{(r^2+a^2)^2H_2\chi}(\beta dt -\alpha\gamma a\sigma) \, , \\
\mathcal{B} &=& \gamma\frac{ms_1s_2}{(r^2+a^2)^2}\Big(\frac{1}{H_1} +\frac{1}{H_2}\Big)
 dt\wedge a\sigma \, , \quad
\mathcal{C} = \gamma\frac{mas_1s_2}{(r^2+a^2)\chi}\sigma\wedge d\sigma \, ,
\eea
where
\bea
\Delta_r &=& g^2(r^2+a^2)^4H_1H_2 +\chi(r^2+a^2)^3
 +\frac{4mag\gamma\alpha\beta}{\chi}(r^2+a^2) \nn \\
&& -\frac{2m(s_1^2\beta^2-s_2^2\alpha^2)}{\chi(s_1^2-s_2^2)}(r^2+a^2)
 +\frac{2ma^2\gamma^2(s_1^2\beta^2-s_2^2\alpha^2)}{s_1^2-s_2^2} \, , \nn
\eea
with the constraint condition
\be
\alpha^2 -\beta^2 = \chi^2(s_1^2-s_2^2) \, . \label{conseq}
\ee
In the above, we have included three extra parameters ($\alpha, \beta, \gamma$) for convenience. But they are subject to one
constraint Eq. (\ref{conseq}), so only two of them are independent, and related to the two charges parameters. They can not be
removed by a coordinate transformation, rather are determined by the requirement that the new form of the solution approach
to all known solutions, as did in Ref. [9] and in the below.

A simple form for the solution is given by choosing
\bea
\alpha = \chi c_1 \, , \quad \beta = \chi c_2\, , \qquad
\gamma = \frac{1}{1+ga} = \frac{1-ga}{\chi}  \, ,
\eea
so we have
\bea
\Delta_r &=& g^2(r^2+a^2)^4H_1H_2 +\chi(r^2+a^2)^3 -2m(1-ga)^2r^2 \nn \\
 && \qquad +4mag(1-ga)(c_1c_2-1)(r^2+a^2) \, . \nn
\eea
When both charges disappear, the metric becomes the vacuum Kerr-AdS$_7$ spacetime
\bea
ds^2 &=& -\frac{1+g^2r^2}{\chi} dt^2 +\frac{(r^2+a^2)^2r^2\, dr^2}{(1+g^2r^2)(r^2+a^2)^3
 -2m(1-ga)^2r^2} \nn \\
&& +\frac{r^2+a^2}{\chi} d\Sigma_2^2 +\frac{r^2+a^2}{\chi}\big(\sigma -g dt\big)^2
 +\frac{2m}{(r^2+a^2)^2}(dt -\gamma a\sigma)^2 \, . \nn
\eea
We hope that the new form for the solution could act as an useful guidance on constructing
the most general solution with two unequal charges and with three unequal rotation parameters
in seven-dimensional gauged supergravity.

\end{document}